\documentclass[9pt, sigconf, screen]{acmart}
\usepackage{amsfonts}
\usepackage{booktabs}
\usepackage{endnotes}
\usepackage{enumitem}
\usepackage{caption}
\usepackage{flushend}
\usepackage{mdwlist}
\setlist{nolistsep}
\setlist[description]{noitemsep,topsep=0pt,parsep=0pt,partopsep=0pt,leftmargin=5pt}
\usepackage[T1]{fontenc}
\usepackage[utf8]{inputenc}
\usepackage[group-separator={,}]{siunitx}
\usepackage{soul}
\usepackage[center,hang,scriptsize,tight]{subfigure}
\usepackage{textcomp}
\usepackage{xspace}

\hypersetup{pdfstartview=FitH,
  pdfpagelayout=SinglePage,
  bookmarksnumbered=true,
  bookmarksopen=true,
  colorlinks=true,
  citecolor=blue,
  linkcolor= blue,
  urlcolor=blue}

\newcommand{\term}[1]{\emph{#1}}
\newcommand{\stress}[1]{\textit{#1}}

\newcommand{\parab}[1]{\vspace{0.08in}\noindent{\bf #1} }
\newcommand{\parai}[1]{\vspace{0.03in}\noindent{\textit{#1}}\quad}

\newcommand{\figcap}[1]{\caption{\textit{#1}}}

\newcommand{\second}[0]{\textsl{second}}

\newcommand{\ums}[1]{\SI{#1}{\milli\second}}
\newcommand{\us}[1]{\SI{#1}{\second}}

\newcommand{\uMbps}[1]{\SI{#1}{Mbps}}

\newcommand{\cut}[1]{}
\newcommand{\iframe}{\stress{I}-Frame\xspace}
\newcommand{\pframe}{\stress{P}-Frame\xspace}
\newcommand{\bframe}{\stress{B}-Frame\xspace}
\newcommand{\bufratio}{\stress{bufRatio}\xspace}
\newcommand{\ratebuf}{\stress{rateBuf}\xspace}

\newcommand{\thesystem}{\emph{ClipStream}\xspace}
\newcommand{\thesystemf}{\emph{ClipStreamFEC}\xspace}
\newcommand{\assim}{\emph{aSSIM}\xspace}

\newlength{\onecolgrid}
\newlength{\twocolgrid}
\newlength{\threecolgrid}
\begin{document}

\fancyhead{}
\settopmatter{printacmref=true, printccs=true, printfolios=true}

\title{The~QUIC Fix for Optimal Video Streaming}

\author{Mirko Palmer, Thorben Krüger, Balakrishnan Chandrasekaran, Anja Feldmann}
\affiliation{
  \institution{Max-Planck-Institut für Informatik\\
    \{mpalmer, tkrueger, balac, anja\}@mpi-inf.mpg.de}
}

\begin{abstract}
Within a few years of its introduction, QUIC has gained traction: a~significant
chunk of traffic is now delivered over QUIC.
The networking community is actively engaged in debating the fairness,
performance, and applicability of QUIC for various use cases, but these debates
are centered around a narrow, common theme:
how does the new reliable transport built on top of UDP fare in different
scenarios?
Evaluation of unreliable delivery in QUIC remains largely unexplored.

The option for delivering content unreliably, as in a best-effort model,
deserves the QUIC designers' and the QUIC community's attention.
We propose extending QUIC to support unreliable streams and discuss a simple use
case of video streaming---an application that dominates the overall Internet
traffic---that can leverage the unreliable streams and potentially bring immense
benefits to network operators and content providers.
We demonstrate, using controlled-environment trials, how to combine
reliable and unreliable streams to outperform TCP and QUIC in video streaming.


\end{abstract}


\copyrightyear{2018}
\acmYear{2018}
\setcopyright{acmcopyright}
\acmConference[EPIQ'18]{Workshop on the Evolution, Performance, and Interoperability of QUIC }{December 4, 2018}{Heraklion, Greece}
\acmBooktitle{Workshop on the Evolution, Performance, and Interoperability of QUIC (EPIQ'18), December 4, 2018, Heraklion, Greece}
\acmPrice{15.00}
\acmDOI{10.1145/3284850.3284857}
\acmISBN{978-1-4503-6082-1/18/12}


\maketitle

\section{Introduction}\label{sec:intro}

Google's Quick UDP Internet Connections (QUIC) protocol offers TCP-like
properties at the application layer on top of
UDP~\cite{Hamilton-InternetDraft2016, Langley-SIGCOMM2017}.
Although the protocol was designed and made public only recently, in 2013, it is
rapidly gaining adoption: nearly $6\%$ of the global Internet traffic flows over
QUIC, and many CDNs and content providers already support
the protocol~\cite{Langley-SIGCOMM2017}; Google, unsurprisingly, leads the Internet in
QUIC adoption and delivers more than $40\%$ of its traffic via
QUIC~\cite{Ruth-PAM2018}. Given the browser support, notably with the Google Chrome
browser even enabling the protocol by default, together with the popularity of
Google's services---the infrastructure of which support QUIC---these adoption
statistics will quickly and significantly increase.

Although QUIC seems to deliver data in a reliable, secure, and fast manner, this
fixation on \stress{only} the reliable-delivery aspect of the protocol (and,
consequently, the lack of support for unreliable delivery) needs a closer
examination.
Naturally, we ask the following questions:
(a) Is the lack of unreliable streams in QUIC really an issue?
(b) Is there a clear use case for a \stress{selectively or partially reliable}
transport, where an application can seamlessly multiplex reliable and unreliable
streams over a single connection?
(c) Is it practical to extend QUIC to offer a partially reliable transport?

To highlight a need to reconsider the strict adherence to reliable transport, we
focus on one class of traffic delivered, today, via QUIC---video streaming.
Video traffic constitutes a significant share of traffic delivered using
QUIC~\cite{Bouzas-WebArticle2018, Langley-SIGCOMM2017}.\footnote{Although this
video traffic over QUIC is only from Google, its YouTube video streaming service
is one of the largest video serving platforms in the Internet.}
The inherent challenges in streaming ``real-time'' video
traffic~\cite{Shenker-JSAC2006, Jiang-CoNEXT2013} over varying, and sometimes
less than ideal, network conditions are only exacerbated by the choice of a
\stress{reliable} transport---so far, TCP.
It is well known that TCP is not suited for video streaming: the rich body of
prior work on optimizing and extending TCP, and adaptive bitrate (ABR) selection
attest to this observation~\cite{Fouladi-NSDI2018, McQuistin-NOSSDAV2016,
Yin-SIGCOMM2015, Huang-SIGCOMM2014, Jiang-CoNEXT2012}.
TCP retransmissions of lost packets in a video stream, inadvertently lead to
\term{stalls} in the video stream. TCP also performs poorly when it encounters
packet losses that are not due to congestion.
By shunning unreliable delivery, QUIC, thus, falls trap to most, if not all, of
TCP's problems for video streaming; in some instances, QUIC has been shown to
perform even worse than TCP for video streaming~\cite{Bhat-NOSSDAV2017}.

The rationale for streaming video via TCP (or, generally, the fixation on
reliable transport), today, is rooted in the economics and feasibility of
streaming infrastructure deployment.
More than $52\%$ of today's Internet traffic is delivered by content delivery
networks (CDNs)~\cite{Cisco-WhitePaper2017}.
When we consider the massive, distributed infrastructure and mature software
stack that CDNs have already deployed for delivering Web traffic, the idea of
streaming video over HTTP, using dynamic adaptive streaming over HTTP (DASH) or
HTTP live streaming (HLS) sounds appealing and practical.
This choice of HTTP, unfortunately, ties video streaming to TCP.
But with CDNs (e.g., Akamai) and popular Web browsers (e.g., Google Chrome)
already supporting QUIC, it is worth revisiting the status quo in video
streaming~\cite{Yakan-WebArticle2018, Ruth-PAM2018, AkamaiBlog-WebArticle2016,
Ponec-Talk2016}.

We share a simple observation to highlight that reliable transports are
ill-suited for video streaming: video data consists of different types of
frames, some types of which do \stress{not} require reliable delivery. The loss
of some types of frames has minimal or no impact (since such losses can be
recovered) on the end-user quality of experience (QoE)~\cite{Feamster-PV2002}.
Therefore, by adding support for unreliable streams in QUIC and offering a
selectively reliable transport, wherein not all video frames are delivered
reliably, we can optimize video streaming and improve end-user experiences.
This approach has several advantages: (a) it builds atop QUIC that is rapidly
gaining adoption; and (b) it involves only a simple, backward compatible,
incrementally deployable extension---support for unreliable streams in QUIC.
These advantages taken together make this approach safe, easy, and practical to
deploy.

We propose a simple extension to QUIC: the addition of unreliable streams.
To demonstrate the benefits of this extension for video streaming and address
the non-trivial challenges of combining both unreliable and reliable transport,
we present \thesystem.~\footnote{\thesystem, our hybrid approach, has no
relation to the online video platform with a nearly identical name.}
Our approach is motivated by a simple observation: not all frames in a video
encoding scheme, such as the widely used H.264, are equally ``important''; some
frames (e.g., {\iframe}s) are more ``important'' than others (e.g., \term{B}-
and {\pframe}s). ``importance'' refers to the implications of the loss of a
frame, contained in a video stream, for the QoE that an end user attributes when
watching that video.

Our streaming solution, \thesystem, thus, uses reliable transport for the
important frames and unreliable transport, for all other frames.
To tackle losses in the unreliable stream, \thesystem uses forward error
correction (FEC), as required.
Supporting such a partially reliable stream, however, introduces other
non-trivial challenges, e.g., synchronization of the streams.
Demonstrating that the partially reliable stream fares well compared to TCP and
QUIC using controlled-environment trials, and addressing the challenges in using
it for video streaming is the central theme of this paper. We summarize our
contributions as follows.

\begin{itemize}[label=$\star$,leftmargin=10pt]
\item We propose the addition of unreliable streams to QUIC. We discuss the ease
  of implementation of this extension and its implications for applications.
  
\item We motivate the extension of QUIC through a simple, practical use case:
  video streaming. 
  To this end, we present \thesystem, a hybrid transport
  protocol that offers selective (or partial) reliability; \thesystem provides
  reliable transport for frames that explicitly request it, and unreliable, best
  effort transport (protected by FEC) for the rest.

\item We present preliminary evaluations---using experiments in controlled
  environments---that show \thesystem outperforms other solutions by a significant
  margin: even under 1.28\% of loss, our approach delivers the video stream
  without compromising video quality, i.e., users see little or no visible
  quality degradation when viewing the video.
\end{itemize}


\section{The Status Quo}\label{sec:motivation}

Streaming video over a reliable transport has remained the status quo for a long
time, but this scheme suffers to sustain a high end-user QoE when the network
conditions are less than ideal.
To highlight some of the problems with current video streaming solutions we
performed a simple experiment where we, in the lab, repeatedly streamed the
``Big Buck Bunny'' video (described in Tab.~\ref{tab:video-clips}) across a
lossy link.
For details on the setup refer~\S\ref{sec:evaluation}.
We varied the loss rates from $0.08$\% to $5.12$\%, and set the link bandwidth
to $\uMbps{20}$ and delay to $\ums{30}$.
We repeated the experiment with several other choices for network parameters
and also using other videos, and observed similar results (not shown).

To assess the performance of a streaming solution, we rely on commonly used,
industry-standard metrics---e.g., \stress{buffering ratio (bufRatio)} and
\stress{rate of buffering (rateBuf)}~\cite{Dobrian-SIGCOMM2011}.
\stress{bufRatio} is defined as the ratio of time spent in re-buffering to the
total video duration, and \stress{rateBuf} is the ratio of the frequency of
re-buffering events to the total number of video frames.
The former captures for each instance of a re-buffering event the duration for
which it lasted and affected end users' experiences, while the latter only
captures the rate at which end users are interrupted in the course of watching a
video.

\begin{figure*}[tp]
    \centering
    \subfigure[Percentage of playback time spent in buffering.]{\label{fig:bufratios-tcp-quic}
      \includegraphics[width=0.32\textwidth,
      keepaspectratio]{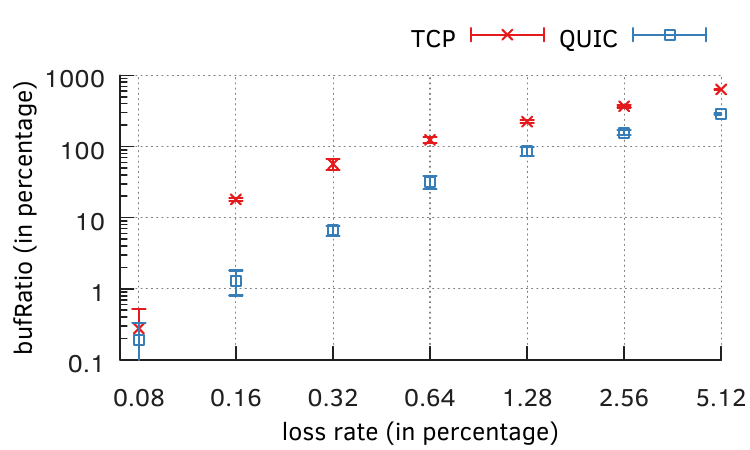}}
    \subfigure[Percentage of frames where stall events occurred.]{\label{fig:ratebuf-tcp-quic}
      \includegraphics[width=0.32\textwidth,
      keepaspectratio]{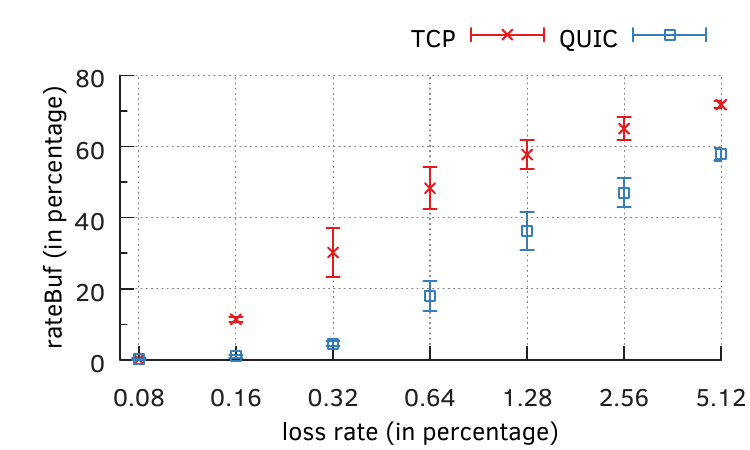}}
    \subfigure[CDF of frame arrival times.]{\label{fig:frametimes-tcp-quic}
      \includegraphics[width=0.32\textwidth,
      keepaspectratio]{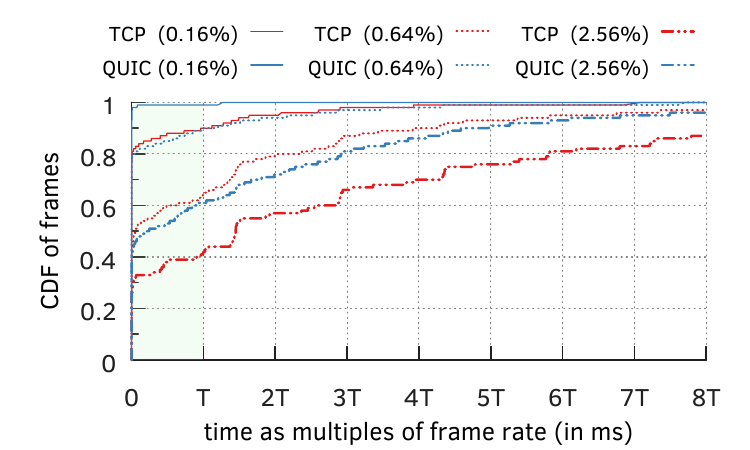}}
    \figcap{TCP and QUIC are not well suited for video streaming. Even at a loss
      rate of $0.64\%$ TCP (QUIC) encounters, in the median, $105\%$ ($30\%$)
      buffering, per Fig. (a), with $50\%$ ($19\%$) of stall events, per Fig.
      (b). At this loss rate, Fig. (c) shows that TCP (QUIC) delivers $64\%$
      ($90\%$) of frames before the deadline (region shaded in green).}
    \label{fig:tcp-quic-status}
\end{figure*}

\parab{TCP}
Despite its shortcomings for streaming videos over the Internet, \emph{TCP} is
still the dominant transport protocol for video streaming, due to the widespread
use of DASH~\cite{Robinson-WebArticle2011}.
The rich body of prior work on optimizing TCP, adaptive bitrate selection
algorithms, or TCP variants highlights TCP's shortcomings~\cite{Fouladi-NSDI2018,
McQuistin-NOSSDAV2016, Yin-SIGCOMM2015, Huang-SIGCOMM2014, Jiang-CoNEXT2012}.
TCP retransmits lost packets without considering if these retransmissions are
``useful'' for the video player; unnecessary retransmissions introduce stalls
and degrade the quality of the video stream. Besides, it is well-known that TCP
performs poorly when it encounters packet losses that are not due to congestion.
Fig.~\ref{fig:bufratios-tcp-quic} shows {\bufratio} as a function of loss,
and, per this figure, even at a loss rate of $0.16\%$---lower than that typically
observed in the Internet~\cite{Sundaresan-SIGCOMM2011}---the video player spends
$20\%$ of the total video time in stalls (i.e., in waiting for the lost packets
to arrive at the playback buffer).
To put this bufRatio in perspective, note that a $1\%$ of {\bufratio} can
reduce user engagement by more than $3$ minutes~\cite{Dobrian-SIGCOMM2011}.
The rate of re-buffering events in Fig.~\ref{fig:ratebuf-tcp-quic} is also
high: at $0.64\%$ loss TCP introduces on average $105\%$ of re-buffering.
A~recent study indicates that traffic policing is highly prevalent world-wide
and induces, globally, an average loss rate of over
$20\%$~\cite{Flach-SIGCOMM2016}: streaming video over TCP under such loss rates
is infeasible.

\parab{QUIC}
Google's QUIC protocol~\cite{Hamilton-InternetDraft2016} takes a positive, albeit
small, step forward towards improving the status quo.
QUIC vastly improves connection establishment times, which might lower the
initial video buffering times, but Ghasemi et al. empirically show that
the impact of throughput on end-users' video quality is higher than that of
latency~\cite{Ghasemi-IMC2016}.
QUIC packs support for better bandwidth estimation and pluggable congestion
control mechanisms, and its transport streams allow applications to seamlessly
multiplex several requests or data exchanges on a single connection to avoid
head-of-line blocking.
The current design, however, demands the use of reliable transport even though, in
principle, unreliable transport options and error correction schemes could be
supported.
Due to this strict adherence to reliable transport, QUIC inherits some of TCP's
issues: Fig.~\ref{fig:frametimes-tcp-quic} shows that even at a loss rate of
$0.64\%$, QUIC fails to deliver $10\%$ of the video frames, i.e., these frames
arrive much later than when they were required, thereby causing stalls. 
Our experiments in a controlled environment show, typically (i.e., in the
median), a relatively high \stress{bufRatio}, in
Fig.~\ref{fig:bufratios-tcp-quic}, and \stress{rateBuf}, in
Fig.~\ref{fig:ratebuf-tcp-quic}, even at a loss rate of 0.64\%.

While ABR schemes help in alleviating some of the issues, they are still akin to
``band-aids'': they are designed to fix transient problems that the underlying
transport fails to handle; besides, switching bitrates has implications for the
end-user QoE~\cite{Garcia-QoMEX2014, Hosfeld-QoMEX2014}.
In case of QUIC, surprisingly, prior work also show that ABR schemes ported to
QUIC operate poorly compared to TCP~\cite{Bhat-NOSSDAV2017}.
Simply switching to UDP for video streaming also does not suffice. The inherent
unreliability of UDP necessitates the use of coding or error-correction
techniques to recover lost packets. \stress{Blindly} coding every packet, in an
application-agnostic manner, to recover from losses poses problems: error
correction schemes have a significant overhead, and unrestricted use of such
schemes even by a small fraction of the users on a network will add significant
load (or traffic) on the network. Besides, without proper congestion control,
the UDP streams will not share network resources equitably with other TCP flows.


\section{A~Primer on Streaming}\label{sec:background}

Today, video streams are being delivered typically via HTTP using
either DASH~\cite{Sodagar-MultiMedia2011} or HLS~\cite{Pantos-RFC2017}.
While both, DASH and HLS, have similar requirements regarding the video format, we
restrict our attention to the codec-agnostic DASH.
When streaming a video via DASH, the client first requests a \term{manifest}
file~\cite{Sodagar-MultiMedia2011}. The manifest specifies the quality levels at
which the video can be delivered, the details of the encoding, and metadata on
the actual video (e.g., name of files and locations) stored on the server.

\begin{figure}[bp]
 \centering
 \includegraphics[width=.6\columnwidth,keepaspectratio]{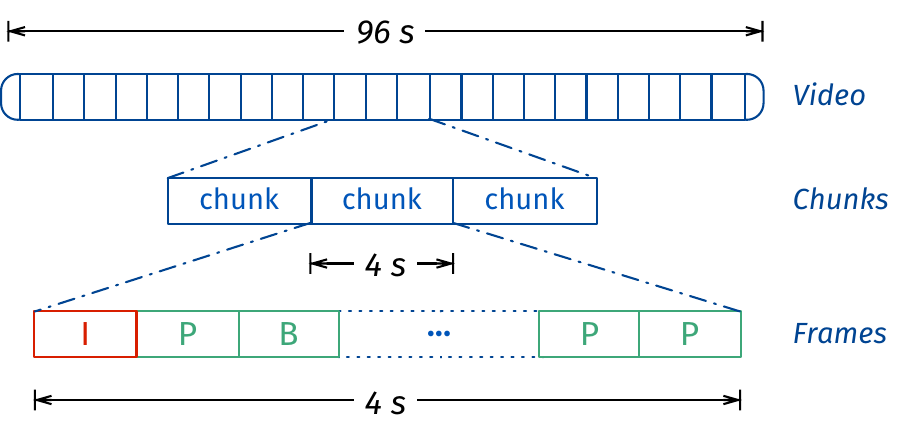}
 \figcap{Components of a video file encoded using the H.264 codec for streaming
   via DASH. Video is split into equally sized chunks, each of which comprises
   one {\iframe} and, depending on the length, several \stress{B-} and
   {\pframe}s.}
 \label{fig:encoding}
\end{figure}

Today, the most widely used video codec in the Internet is
H.264~\cite{encoding.com-WhitePaper2018}. To encode a video using H.264 and
stream it via DASH, the video data is split into \term{chunks}, each of which
contain the same number of video frames,\footnote{Except, perhaps, the last
chunk, which might contain fewer frames.} as illustrated in
Fig.~\ref{fig:encoding}.
Often the video is encoded at different qualities (i.e., at different bitrates
and/or resolutions) to enable ABR switching at the receiver or video player; in
case of congestion, for instance, the video player might fetch the next chunk at
a lower quality and avoid stalling the video stream. To allow fast switching,
the chunk duration is commonly in the range of $\us{1}$ to $\us{10}$.

\begin{sloppypar}
The H.264 codec defines three types of \term{slices}: \term{I-}, \term{P-}, and
\term{B-slices}~\cite{Juurlink-SpringerBriefs2012}.
We simplify, however, the H.264 specification's terminology in that we
do not use the term \term{slice} explicitly.
Each frame, in our terminology, consists of only one (\term{I-}, \term{P-}, or
\term{B-}) slice; an {\iframe}, for instance, refers to a frame consisting
of only H.264 \term{I-slices}.
\end{sloppypar}

To help video players \stress{instantly} start playback upon receiving a chunk
(or after buffering enough chunks), each chunk needs to start with an {\iframe}.
Since {\iframe}s are \stress{independent} frames, they can be rendered
instantaneously. This lack of dependence on other frames results in the
{\iframe}s being significantly large in size, and, hence, they should be used
sparingly to keep the size of the video file small. To seamlessly switch between
the different quality levels, we need, however, an {\iframe} at the start of
each chunk. In contrast, {\pframe} depends on one or more previous frames, which
can be of any type, and {\bframe}s depend on both previous as well as following
frames. These inter-dependencies confirm a simple observation: {\iframe}s~are
essential and, therefore, should be well protected against loss while
\stress{P-} and {\bframe}s are less essential~\cite{Feamster-PV2002,
Albanese-TOIT1996, Shimamura-SPIE1988, Yang-TOMM2005}.


\section{The QUIC fix}\label{sec:quic-fix}

The design of an optimal transport for video streaming hinges on two simple
observations: \stress{(a) I-Frames should be reliably streamed}, and \stress{(b)
  It is relatively easy to recover from B- and P-Frame losses}.

We \stress{require} an {\iframe} to start video playback, and, hence, this frame
should be reliably delivered; the playback of the remaining frames (of the
concerned chunk) depend on it. Since the remaining frames encode only the
\stress{deltas} or differences with reference to the starting {\iframe}, the
loss of the {\iframe} renders the deltas of no use, resulting in significant
implications for the QoE.
Regarding losses, a~recent study~\cite{Orosz-IFIP2015} shows that the impact of
\stress{B-} and {\pframe} losses on end-user QoE is less severe than that of
{\iframe} losses.
In DASH streaming, we can quickly recover from losses after each chunk, which is
at most a few seconds long, if we transfer the {\iframe} of each chunk
\stress{reliably}. If sufficient {\iframe}s are available (at brief-enough
intervals) the impact of QoE should \stress{not} be significant, despite losses
in other frames. In practice, we can also use forward error correction (FEC)
mechanisms, while carefully measuring the overheads introduced, to correct for
losses in \stress{B-} and {\pframe}s.

QUIC offers a good starting point for redesigning video transport.
QUIC supports multiple streams within a single association and decouples
congestion control from retransmission.
In particular, QUIC's congestion control and acknowledgments operate on a
per-packet basis while retransmissions are realized on a per-stream basis.
This feature enables the sender to selectively retransmit
or to introduce FEC on a per-stream basis, and, thus, allows, in principle,
reliable and unreliable streams within the same association.
Extending QUIC to support such selective delivery of the video frames over
either reliable or unreliable streams, as required based on the frame type,
introduces several non-trivial challenges.

\stress{$\bullet$ Adding unreliable streams to QUIC.}
Streams in QUIC offer a light-weight, in-order byte-stream
abstraction~\cite{Iyengar-InternetDraft2018}; they are individually
flow-controlled and subject to congestion control.
Streams, however, only offer reliable delivery.
Indeed, QUIC makes, quoting the current IETF Internet
draft~\cite{Iyengar-InternetDraft2018}, \stress{``no specific allowance for
partial reliability. Endpoints MUST be able to deliver stream data to an
application as an ordered byte-stream.''}
This limitation makes it challenging to add support for unreliable streams and
ensuring such changes are backward compatible, i.e., do not break QUIC's flow
control and congestion control logic.
We exploit a simple insight to solve this problem: to support unreliable streams
we need to change \stress{only} the way retransmissions are handled.
More concretely, at the sender, we choose to replace retransmission of missed
data with \stress{opportunistic} transmission of the next byte range, i.e., the
set of next QUIC frames. At the receiver we do not change the acknowledgment
strategy: all packets, including out-of-order packets, are acknowledged using
selective ACKs. The sender, hence, receives the feedback on lost packets to
adjust its congestion window, but it sends \stress{new} rather than the lost
data. This approach ensures that transmission can continue without breaking flow
or congestion control.
We also leverage the existing re-order buffer at the receiver: an out-of-order
packet is inserted into the byte-stream within this buffer unless the data has
already been consumed by the application. If the application tries to consume
``missing'' byte-ranges the byte-stream is filled with zeros.

\stress{$\bullet$ Negotiating appropriate streams.}
The choice of reliable as well as unreliable QUIC streams leads to an obvious
follow-up question: what data should be delivered reliably?
Based on prior work on the impact of losses of different types of frames on
video quality~\cite{Orosz-IFIP2015, Feamster-PV2002}, we deliver {\iframe}s over
a reliable stream and the other kinds ---\stress{B-} and {\pframe}s---over unreliable streams.
Since unreliable streams are initiated (or requested) by the client, we 
reuse the QUIC handshake mechanism, which includes the capabilities
of the sender or receiver, to advertise and negotiate support for unreliable
streams.

\stress{$\bullet$ Selectively enabling reliability.}
We can either provide a \stress{meta} stream within QUIC that dictates how to
selectively offer reliability---by tagging individual QUIC frames as reliable or
unreliable---or implement an interface in QUIC that facilitates a client (e.g.,
Web browser or video player) in opening reliable as well as unreliable streams.
In either case video frames are sent via the appropriate streams based on
application-offered insights into reliability. The receiver may also use this
meta-information to de-multiplex the streams and deliver the data to the
application.

\stress{$\bullet$ Synchronizing partially reliable QUIC streams.}
Multiplexing the video frames over reliable and unreliable streams introduces
another challenge: \stress{How will the receiver (of a video stream) combine the
frames from the different streams into the appropriate order in the playback
buffer?}
To this end, we can add a reliable \stress{control stream} to signal
multiplexing and demultiplexing information for the different streams to the
client. The control stream helps the client to orchestrate its reads from the
different streams and, thus, re-assemble the video file.
Lastly, an issue that arises in case of out-of-order delivery is that the
receiver will be unable to determine the end of the transmission on a stream;
suppose, for instance, that the last (QUIC) frame is lost.
To cope with this issue, one solution, which we choose, is to reliably transfer
end-of-stream markers.

\stress{$\bullet$ Tagging each video frame with reliability markers.}
The sender of the video stream needs to tag each frame as reliable or unreliable
to deliver it via the appropriate stream. Naturally, the sender has to parse and
decode the video file, and mark each video frame to indicate whether it requires
reliability. \stress{Is it feasible for the sender to decode and tag frames?}
This need to decode the video, in contrast to treating it as an opaque object,
induces some overhead, but it is either a one-time cost or incurs only a small
overhead.
Indeed, to enable the widely used industry practice of supporting multiple
resolutions as well as bitrate selections by clients, e.g., via
DASH~\cite{Sodagar-MultiMedia2011}, video files are typically encoded \stress{a
priori} at different (predefined) resolutions. In case of live streams, videos
are transcoded on demand. The tagging or reliability information (i.e., marking
of frames), in either case, can be seamlessly integrated into this encoding
process.
The only remaining overhead is that the server must parse these
reliability tags to choose the appropriate QUIC stream.
We can, however, add these reliability tags to the DASH manifest files allowing
clients to initiate the appropriate streams and deliver data corresponding to
each without any additional overhead.


\begin{table}[tp]
{\small
  \centering
	\caption[]{Video file characteristics: Resolution (Res); Bitrate, in Mbps
      (Br); Duration (Dur); Size, in MB; \#I-Frames (\#I); and \#B-/P-Frames
      (\#B/P).\protect\footnotemark}
    \label{tab:video-clips}
    \begin{tabular}{r|rr|rr|rr}
      \toprule
      \textit{Video} & \textit{Res} & \textit{Br} & \textit{Dur} & \textit{Size} & \textit{\#I} & \textit{\#B/P}\\
      \midrule
      Big Buck Bunny & 1080p & 5 & 296.21 & 176 & 75 (1\%) & 7,031\\
      Sintel & 1080p & 5 & 296.21 & 182 & 75 (1\%) & 7,031 \\
      Tears of Steel & 1080p & 5 & 296.21 & 182 & 75 (1\%) & 7,031 \\
      \bottomrule
    \end{tabular}
  }
\end{table}

\footnotetext{We shortened the videos to have the same length: The number of
  {\iframe}s (one per 96 frames or 4 s) and the combined number of
  \stress{P-} and {\bframe}s is the same across the videos. We do not discuss
  control frames, they account for $0.05 \%$ of the video.}

\subsection{Prototype}\label{sec:prototype}

\begin{figure*}[t!]
  \centering
  \subfigure[Playback time spent in buffering.]{\label{fig:in-lab-bufratio}
    \includegraphics[width=0.32\textwidth,
    keepaspectratio]{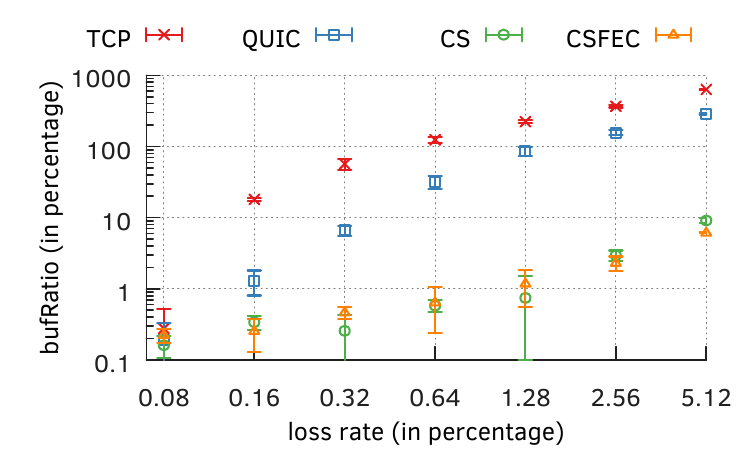}}
  \subfigure[Relative frequency of stall events.]{\label{fig:in-lab-ratebuf}
    \includegraphics[width=0.32\textwidth,
    keepaspectratio]{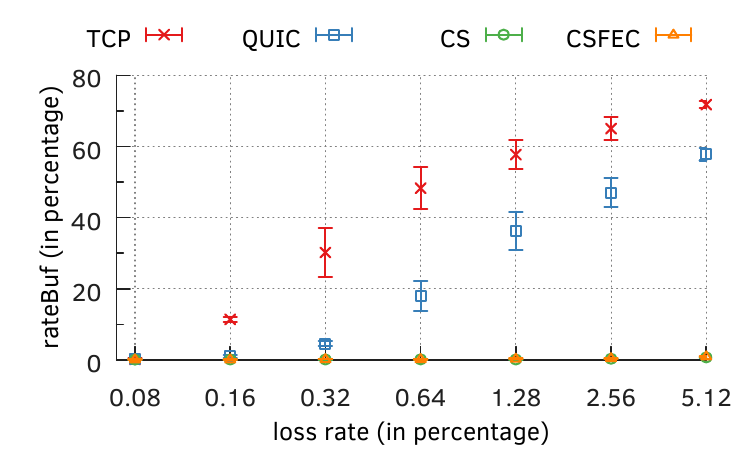}}
  \subfigure[Adjusted SSIM of reference \& received frames.]{\label{fig:in-lab-ssim}
    \includegraphics[width=0.32\textwidth,
    keepaspectratio]{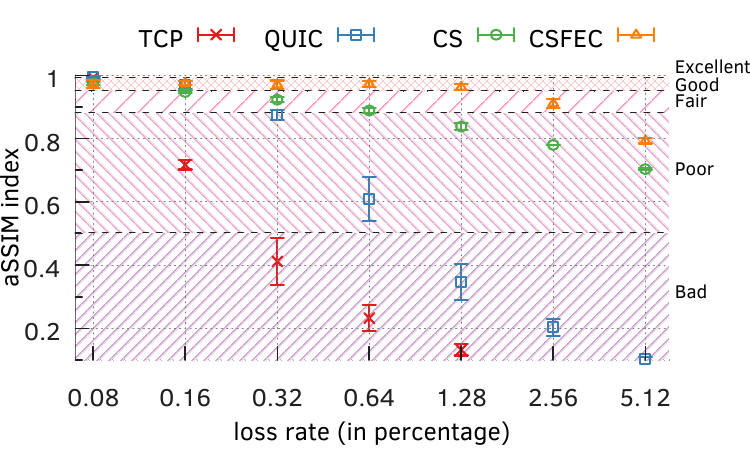}}
  \figcap{{\em\thesystem} (CS) and {\em \thesystemf} (CSFEC) outperform TCP and
    QUIC across a wide range of loss values.}
  \label{fig:in-lab-perf}
\end{figure*}

We developed our prototype based on \texttt{quic-go}~\cite{quic-go},
specifically the version with commit ID \stress{c852814} from Oct. 2017. The
implementation of \thesystem comprises a \stress{shim} layer application and our
modifications to QUIC. The latter involves only \textasciitilde\stress{200}
lines of code. We realized unreliable streams by instructing the server (or
sender) to avoid retransmissions in case of loss. We implemented a new interface
for unreliable streams for allowing the clients (e.g., video player, or Web
browser) to explicitly specify the required stream type. In addition, we added a
reliable \stress{control stream} to signal multiplexing and demultiplexing
information for the different streams to the client so that it can orchestrate
its reads from the streams and, thus, re-assemble the video. The shim processes
this control stream: on the server side, it receives \stress{untagged} video
files and marks the video frames, as reliable or not, on the fly; on the client
side, it reassembles the video frames before feeding it to the video player.

When the client attempts to consume data that has not yet arrived or that has
been lost, it will receive a buffer of zeros;\footnote{Modern video players,
e.g., VLC, are capable of decoding zero-padded streams without any issue.} the
buffer is sized to match the missing QUIC frames.  We currently transmit the
last byte of each frame reliably.
The shim compensates for some of the loss, in unreliable streams, by using
Reed-Solomon erasure coding technique (an implementation of which is available
as a Go library~\cite{reed-solomon-go}) for each of the video frames. This FEC
coding scheme is well-suited for our needs as it can deal with various kinds of
byte errors including bursts.
We configured the FEC scheme to deliver each video frame with an overhead of 1/3
of the frame size as redundant data; more concretely, we split each video frame
into 18 shards, compute 6 parity shards, and deliver the 24 shards.

%

\section{Evaluation}\label{sec:evaluation}

To compare and contrast the performance of \thesystem (and \thesystemf, which
adds FEC support) with QUIC~\cite{Iyengar-InternetDraft2018} and TCP, we
streamed videos from one host to another through an intermediate host, called
the \stress{shaper}.
The hosts are physical machines running Debian Linux (version 9) with kernel
version \textit{4.9.91.1}.
We used the \texttt{tc} utility in Linux for emulating specific link capacities
and delays.
We set the link capacities to $\uMbps{20}$, which is large enough to accommodate
the video streams and FEC overheads, and we sized buffers to hold $1000$
packets, chiefly to accommodate QUIC's
burstiness~\cite{IETF93-QUIC-BarBoF-Talk2015}.
To emulate typical ``last mile'' latencies, we configured a $\ums{30}$ delay on
the link between the client and shaper.
Lastly, we captured packet traces using \texttt{tcpdump} and instrumented the
server-side and client-side video streaming software for obtaining frame-level
timing data.

\parab{Data Set.}
We selected videos (Tab.~\ref{tab:video-clips}) that are
deemed standard \cite{dash-dataset} and widely used in the literature: ``Big
Buck Bunny'', for instance, was used in \cite{Lederer-2012}, ``Sintel'' in
\cite{Zia-MMSys2016}, and ``Tears of Steel'' in~\cite{Hosfeld-QoMEX2014}.
We re-encoded these videos using the \texttt{ffmpeg} utility to adhere to a
frame rate of $24$ fps.
To simplify evaluation, the original videos were cut to be of uniform length
spanning $\us{296.21}$.
The videos require a \stress{minimum} bandwidth ($B_{min}$) of approximately
$\uMbps{5}$, and nearly $1\%$ of the frames in the video file are {\iframe}s.

\parab{SSIM \& \assim.}
In addition to \bufratio and \ratebuf, we compute the \term{structural
similarity (SSIM)}~\cite{Wang-TIP2004} index values to objectively estimate the
stream quality.
SSIM index looks at the quality of the received frames, but ignores the time at
which the frames were delivered.
When a video frame arrives after its deadline the client encounters a stall,
significantly degrading the perceived quality of the video.
SSIM is, hence, \stress{not a good metric} for evaluating either TCP or QUIC.
To capture the effect of these stalls, we compute an \stress{adjusted} SSIM
(\assim) score wherein each frame period (i.e., $1/f_T$, where $f_T$ is the
duration or time span of a frame) over the duration of the stall is assigned an
SSIM index of zero.
In assigning these \assim scores, we are still being generous in the evaluation
of the reliable transports: we assume that despite the stalls the end user will
watch the video rather than abandoning the stream, which seems to be the norm
according to prior work~\cite{Hossfeld-QoMEX2012, Ghadiyaram-GlobalSIP2014,
Dobrian-SIGCOMM2011, Garcia-QoMEX2014, Zhao-CommSurveysTuts2016}.
To estimate the subjective video quality, we map the \assim values to Mean
Opinion Score (MOS) values (based on \cite{Zinner-QoMEX2010}); the MOS values,
e.g., ``excellent'', ``good'', and ``bad'', reflect the subjective measure of
quality perceived by the user.

\parab{Controlled-Environment Trials.}
We streamed the video files under different loss rates, repeating $10$ times
for each loss rate. We computed the mean, median, and standard deviations of the
three performance metrics, \bufratio, \ratebuf, and \assim.
We repeated the experiments with several combinations of the network
parameters---bandwidth, buffer size, and delays; we omit some of the plots in the
interest of space, but discuss the relevant results in the text.
Since prior work shows that switching between quality levels has a negative
impact on QoE~\cite{Garcia-QoMEX2014, Hosfeld-QoMEX2014}, we only use a single
quality level in our experiments. The evaluations, hence, show the ability of
\thesystem to sustain the same quality level under varying loss rates; more
quality levels allow \thesystem more freedom (although each switch affects
QoE), and we leave evaluation with multiple quality levels to future work.

Under no loss, {\bufratio} and {\ratebuf} for all four transport protocols is
rather small---less than $0.25\%$. Overall, TCP was the worst protocol for
both metrics, and both \thesystemf and \thesystem outperform QUIC.
Per Fig.~\ref{fig:in-lab-bufratio} and~\ref{fig:in-lab-ratebuf} we observe
that the {\bufratio} and {\ratebuf} for both \thesystem and \thesystemf
(abbreviated as CS and CSFEC, respectively, in the figures) absolutely dominate
that of TCP and QUIC.
The \ratebuf values for both \thesystem and \thesystemf are very close to $0$\%,
with the maximum being $0.012$\%.
These low {\ratebuf}s are due to \thesystem streaming only a small percentage
(approx. $1\%$ by count or $12\%$ by size) of the overall video stream reliably;
the potential for stalls, hence, is rather small. \thesystem, hence, imposes
the \stress{bare minimum} load, even at loss rates as high as $5.12$\%.

The plot of \assim values as a function of loss rate, in
Fig.~\ref{fig:in-lab-ssim}, also shows that \thesystem performs better than
the rest.
The QoE for TCP drops very quickly from ``excellent'' to ``bad''; even
at a low loss rate of $0.32$\%, TCP delivers a typical \assim value that is less
than $0.5$, far below what is typically considered ``acceptable'' quality.
The QoE for QUIC stays above ``fair'' quality for loss rates smaller than
$0.32$\%, but drops to ``bad'' above $1$\% loss.
\thesystem sustains ``fair'' quality video until $0.64$\% loss and does not
reach ``bad'' quality even at $5.12$\% loss.
\thesystemf significantly improves upon \thesystem, owing to the use of FEC,
delivering ``good'' quality till $1.28$\% and ``fair'' until $2.56$\%.



\section{Related Work}\label{sec:related}

There exists a large body of prior work on video streaming.
Several studies have, for instance, looked at factors affecting
QoE~\cite{Ghasemi-IMC2016, Dobrian-SIGCOMM2011} and on designing optimal
streaming infrastructures~\cite{Liu-SIGCOMM02012, Jiang-CoNEXT2013}.
In this section we briefly discuss only those most relevant to our work.

\begin{sloppypar}
\parai{Adaptive bitrate schemes.}
Buffer-based and rate-based schemes that dynamically adapt the video
bitrate~\cite{Mao-SIGCOMM2017, Sun-SIGCOMM2016, Yin-SIGCOMM2015,
Huang-SIGCOMM2014, Jiang-CoNEXT2012} suffer invariably from the limitations of
the underlying transport: these schemes simply operate on top of an existing
transport protocol that does not discriminate between the different types of
frames in the video stream.
While they help in improving end-user QoE, simply porting over ABR to QUIC
offers poor performance~\cite{Bhat-NOSSDAV2017}.
\end{sloppypar}

\begin{sloppypar}
\parai{TCP variants \& ``tweaks''.}
TCP variants such as TCP-RTM~\cite{Liang02tcp-rtm} and TL-TCP
~\cite{Mukherjee-ICNP2000} either ignore retransmissions or avoid retransmitting
data that have already missed the deadline. The former needs support for loss
recovery to be built into the application and the latter requires application's
cooperation to obtain the deadlines: both complicate application design, making
deployment impractical, if not impossible.
Brosh et al.~\cite{Brosh-SIGMETRICS2008} suggest optimizations to make TCP more
friendly for delivering real-time media. In a similar vein, Goel et
al.~\cite{Goel-TOMM2008} tune TCP's send buffer for mitigating delays. While
these optimizations are important, they will be even more beneficial when
applied selectively to only the portion of data that requires reliability in the
first place.
\end{sloppypar}

\parai{Partial reliability.}
McQuistin et al.~\cite{McQuistin-NOSSDAV2016} propose a novel TCP variant that
uses retransmissions to deliver new data, instead of the lost data. The idea of
using the retransmissions to send new data alleviates some but not all of the
overhead; \stress{B-} and {\pframe}s that have not missed their deadlines will
still be retransmitted.
\cite{Feamster-PV2002} explores the effect of selective reliability for
streaming MPEG-4 video via RTP, necessitating substantial
changes to the network stack.
\thesystem requires minimal changes and can be deployed incrementally.

\parai{Error-correction schemes.}
Kim et al.~\cite{Kim-CoRR2012} propose CTCP, which codes data in an
application-agnostic manner, to improve performance in lossy channels. CTCP's
indiscriminate coding of all video frames by a significant number of users
might, under certain conditions, overwhelm the network capacity. \thesystem can
benefit, however, from using CTCP's adaptive coding scheme for delivering
\stress{B-} and {\pframe}s.


\section{Summary \& Outlook}\label{sec:conclusion}

The increasing adoption of QUIC on the server side (e.g., CDNs) as well as the
client side (e.g., Google Chrome browser) offers us the unprecedented
opportunity to rethink about an ideal transport protocol for video streaming.
We show that such an ideal transport, exploiting partial reliability, can be
realized simply through the addition of unreliable streams to QUIC.
We already submitted a draft to the QUIC Working
Group~\cite{tiesel-quic-unreliable-streams-01} to add support for unreliable
streams, and plan on following up with insights and observations from our
experience of implementing unreliable streams in QUIC and leveraging it in
\thesystem for use in video streaming.
While our preliminary evaluation of the selective use of reliability for video
streaming shows our approach to be better than TCP and QUIC, we envision conducting
real-world experiments (i.e., over the Internet) and comparing our approach with
ABR schemes.


\bibliographystyle{abbrv}
\bibliography{references}

\end{document}
